\documentclass[seceq]{ptptex}

\usepackage{graphicx}

\title{
Double-$\Lambda$ hypernuclei in the relativistic mean-field theory
}

\author{
Hong {\sc Shen}$^{1,}$\footnote{E-mail address: songtc@nankai.edu.cn},
Fang {\sc Yang}$^{1,}$\footnote{E-mail address: yangfang022@mail.nankai.edu.cn}
and Hiroshi {\sc Toki}$^{2,}$\footnote{E-mail address: toki@rcnp.osaka-u.ac.jp}
}

\inst{
$^1$Department of physics, Nankai University, Tianjin 300071, China\\
$^2$Research Center for Nuclear Physics (RCNP), Osaka University,\\
Ibaraki, Osaka 567-0047, Japan
}

\recdate{
}

\abst{
We study the properties of double-$\Lambda$ hypernuclei in the
relativistic mean-field theory, which has been successfully used
for the description of stable and unstable nuclei. With the
meson-hyperon couplings determined by the experimental binding
energies of single-$\Lambda$ hypernuclei, we present a
self-consistent calculation of double-$\Lambda$ hypernuclei in the
relativistic mean-field theory, and discuss the influence of
hyperons on the nuclear core. The contribution of two mesons with
dominant strange quark components (scalar $\sigma^*$ and vector
$\phi$) to the $\Lambda\Lambda$ binding energy of double-$\Lambda$
hypernuclei is examined.
}

\begin{document}

\maketitle

\section{Introduction}
\label{introduction}

The study of hypernuclei has been attracting great interest of nuclear physicists
due to its important role in providing information about hyperon-nucleon and
hyperon-hyperon interactions. Such information is crucial for
understanding the properties of multi-strange systems and neutron
stars~\cite{ann94,prc96s,prc99,jpg02s,prc04s,prc00s}.
The most extensively studied hypernuclear system is the single-$\Lambda$
hypernucleus which consists of a $\Lambda$ particle coupled to the nuclear
core. There exist many experimental data for various single-$\Lambda$
hypernuclei over almost the whole mass table~\cite{npa88,prl91,prc96}.
However, only three double-$\Lambda$ hypernuclei,
$^{6}_{\Lambda\Lambda}\rm{He}$, $^{10}_{\Lambda\Lambda}\rm{Be}$, and
$^{13}_{\Lambda\Lambda}\rm{B}$, have been identified
experimentally~\cite{2lexp1,2lexp2,2lexp3,2lexp4}.
A recent observation of the double-$\Lambda$ hypernucleus
$^{6}_{\Lambda\Lambda}\rm{He}$, called the Nagara event~\cite{2lexp5},
has had a significant impact on strangeness nuclear physics.
The Nagara event provides unambiguous identification of
$^{6}_{\Lambda\Lambda}\rm{He}$ production
with precise $\Lambda\Lambda$ binding energy value
$B_{\Lambda\Lambda}=7.25 \pm 0.19^{+0.18}_{-0.11}\;\rm{MeV}$,
which suggests that the effective $\Lambda\Lambda$ interaction should be
considerably weaker ($\triangle B_{\Lambda\Lambda}=B_{\Lambda\Lambda}
(^{6}_{\Lambda\Lambda}\rm{He}) - 2B_{\Lambda}(^{5}_{\Lambda}\rm{He})
\approx 1\;\rm{MeV}$) than that deduced from the earlier measurement
($\triangle B_{\Lambda\Lambda}\approx 5\;\rm{MeV}$)~\cite{2lexp2}.
The weak $\Lambda\Lambda$ interaction suggested by the Nagara event
has triggered great interest in theoretical studies of double-$\Lambda$
hypernuclei by using various approaches, such as cluster models,
Faddeev calculations, and coupled channel methods~\cite{2lth1,2lth2,2lth3,2lth4}.
It has also been used to examine the properties of strange hadronic
matter~\cite{prc04s,jpg04s,prc03s1,prc03s2}.

The purpose of this paper is to present a self-consistent calculation
of double-$\Lambda$ hypernuclei in a wide range of mass number $A$
within the framework of relativistic mean-field theory (RMF),
and also to examine possible contributions of the two strange mesons
$\sigma^*$ and $\phi$ to the $\Lambda\Lambda$ binding energy.
In principle, exact few-body calculations are more appropriate
for light hypernuclei, but they are not available for heavy systems.
Therefore, we adopt the RMF theory to study the mass number dependence
of various quantities in double-$\Lambda$ hypernuclei.
By performing the self-consistent calculation in the RMF theory,
we hope to constrain the meson-hyperon couplings by reproducing
available experimental data, especially the Nagara data,
which suggests a weak $\Lambda\Lambda$ interaction. It is well known
that these coupling constants play important roles in multi-strange
systems and neutron stars~\cite{prc96s,prc99,jpg02s,prc04s,prc00s,
jpg04s,prc03s1,prc03s2,prl93}.

The RMF theory has been quite successfully used for the description
of nuclear matter and finite nuclei, including unstable nuclei and
deformed nuclei~\cite{sw86,grt90,hs96}. It has also been applied to
predict the equation of state of dense matter for the use in
supernovae and neutron stars~\cite{prc99,npa98}.
There are many works using the RMF theory for the study of single-$\Lambda$
hypernuclei~\cite{1lth1,1lth2,1lth3,1lth4,1lth5,1lth6}, in which
the meson-hyperon couplings were determined by fitting the
experimental $\Lambda$ binding energies of single-$\Lambda$ hypernuclei.
The $\Lambda$ binding energy is obtained experimentally by
$B_{\Lambda}\left (^{A}_{\Lambda}Z \right)=
 B\left (^{A}_{\Lambda}Z \right)-B\left (^{A-1}Z \right)\;$~\cite{prc96},
which might be different from the $\Lambda$ single-particle
energy due to the core polarization effect. It has been shown that
the rearrangement energy, which represents the core polarization effect,
decreases with increasing mass number $A$,
and is negligible in most cases~\cite{1lth1,prc98s}.
Therefore, the experimental $\Lambda$ binding energy is usually
considered as the $\Lambda$ single-particle energy, and is used
to be compared with the calculated $\Lambda$ single-particle energy
in various models. In the present work, we determine the meson-hyperon
couplings by reproducing the experimental $\Lambda$ binding energies
of single-$\Lambda$ hypernuclei, then apply the RMF model to study
the properties of double-$\Lambda$ hypernuclei. We discuss also the
core polarization effect in double-$\Lambda$ hypernuclei.
For hypernuclear systems containing more than one hyperon,
it has been discussed in early works~\cite{prl93,2lth5} that there might
be contributions from the two mesons with dominant strange quark components,
scalar $\sigma^*$ and vector $\phi$, which couple exclusively to hyperons.
We examine these contributions to the $\Lambda\Lambda$ binding energies
of double-$\Lambda$ hypernuclei.
In Ref.~\citen{2lth5}, the $\Lambda\Lambda$ correlation effect has been
estimated by solving a Schrodinger equation with spherical harmonic
oscillator potentials provided by the nuclear core and an effective
$\Lambda\Lambda$ interaction. However, the approximations used in their
treatment have violated the self-consistency between the potentials
and the baryon wave functions. In the present work, we would rather
keep the self-consistency of the calculation than take into account
the $\Lambda\Lambda$ correlation seriously.

The outline of this paper is as follows. In Sec.~\ref{sec:model}
we briefly explain the formulation of hypernuclear systems in the RMF
theory. The model parameters and the method to determine the
meson-hyperon couplings are discussed in Sec.~\ref{sec:parameters}.
The calculated results of double-$\Lambda$ hypernuclei
are presented in Sec.~\ref{sec:results}.
Sec.~\ref{sec:summary} is devoted to a summary.

\section{Model}
\label{sec:model}

In the RMF theory, baryons interact via the exchange of mesons.
The baryons involved in the present work are nucleons and
$\Lambda$ hyperons, while the exchanged mesons consist of
isoscalar scalar and vector mesons ($\sigma$ and $\omega$) and
isovector vector meson ($\rho$). When it is applied to a strange
nuclear system containing more than one hyperon, such as a
double-$\Lambda$ hypernucleus, we consider two models. The first
one (model 1) is a simple application of the RMF theory
containing only usual mesons ($\sigma$, $\omega$, and $\rho$),
and the second one (model 2) incorporates two additional mesons
(scalar $\sigma^*$ and vector $\phi$) which couple exclusively to
hyperons. The effective Lagrangian, after the mean-field
approximation is applied, may be written as
\begin{eqnarray}
\label{eq:lag}
{\cal L} &=&
\bar\psi\left[ i\gamma_\mu\partial^\mu-M_N
-g_\sigma \sigma
-g_\omega \gamma^0 \omega
-g_\rho   \gamma^0 \tau_3 \rho
-e        \gamma^0 \frac{1+\tau_3}{2} A
\right] \psi  \\ \nonumber
 & &
+\bar\psi_{\Lambda} \left[ i\gamma_\mu\partial^\mu-M_{\Lambda}
-g^{\Lambda}_\sigma     \sigma
-g^{\Lambda}_{\sigma^*} \sigma^*
-g^{\Lambda}_\omega     \gamma^0 \omega
-g^{\Lambda}_\phi       \gamma^0 \phi
+\frac{f^{\Lambda}_{\omega}}{2M_{\Lambda}} \sigma^{0i} \partial_i \omega
\right] \psi_{\Lambda} \\ \nonumber
 & &
-\frac{1}{2} (\bigtriangledown\sigma)^2
-\frac{1}{2} m_\sigma^2\sigma^2
-\frac{1}{3} g_2\sigma^3-\frac{1}{4} g_3\sigma^4
-\frac{1}{2} (\bigtriangledown\sigma^*)^2
-\frac{1}{2} m_{\sigma^*}^2 {\sigma^*}^2  \\ \nonumber
 & &
+\frac{1}{2} (\bigtriangledown\omega)^2
+\frac{1}{2} m_\omega^2\omega^2
+\frac{1}{4} c_3\omega^4
+\frac{1}{2} (\bigtriangledown\phi)^2
+\frac{1}{2} m_\phi^2\phi^2, \\ \nonumber
 & &
+\frac{1}{2} (\bigtriangledown\rho)^2
+\frac{1}{2} m_\rho^2\rho^2
+\frac{1}{2}(\bigtriangledown A)^2 ,
\end{eqnarray}
where $\psi$ and $\psi_{\Lambda}$ are the Dirac spinors for
nucleons and $\Lambda$ hyperons. The mean-field values of the
exchanged mesons are denoted by $\sigma$, $\omega$, $\rho$,
$\sigma^*$, and $\phi$, respectively.
We note that $\sigma^*$ and $\phi$ have no contribution to
a single-$\Lambda$ hypernucleus because these mesons are
assumed to be exchanged exclusively between two hyperons
due to the Okubo-Zweig-Iizuka (OZI) rule. $A$ is the electromagnetic
field which couples only to the protons. The $\Lambda$ hyperon is
a charge neutral and isoscalar particle so that it does not couple
to $\rho$ and $A$. We include the non-linear terms for $\sigma$
and $\omega$ mesons, which are introduced to reproduce
properties of finite nuclei and the feature of the
nucleon self-energy obtained in the relativistic
Bruckner-Hartree-Fock theory~\cite{tm1}.
We taken into account the tensor coupling between $\omega$ and
$\Lambda$ as suggested in previous works~\cite{1lth2,1lth3,1lth4,1lth5}.
It is known that the tensor coupling is capable of resolving the
problem of the small spin-orbit interaction in single-$\Lambda$
hypernuclei, but it has negligible influence on the $\Lambda\Lambda$
binding energy of double-$\Lambda$ hypernuclei~\cite{2lth6}.

The Dirac equations for nucleons and $\Lambda$ hyperons have the
following form:
\begin{eqnarray}
\label{eq:driac1}
 & & \left[
i\gamma_{\mu}\partial^{\mu}-M_N
-g_\sigma \sigma
-g_\omega \gamma^0 \omega
-g_\rho   \gamma^0 \tau_3 \rho
-e        \gamma^0 \frac{1+\tau_3}{2} A
\right]\psi
= 0,
\\
\label{eq:driac2}
 & & \left[
i\gamma_{\mu}\partial^{\mu}-M_{\Lambda}
-g^{\Lambda}_\sigma     \sigma
-g^{\Lambda}_{\sigma^*} \sigma^*
-g^{\Lambda}_\omega     \gamma^0 \omega
-g^{\Lambda}_\phi       \gamma^0 \phi
+\frac{f^{\Lambda}_{\omega}}{2M_{\Lambda}} \sigma^{0i} \partial_i \omega
\right]\psi_\Lambda
= 0.
\end{eqnarray}
The equations of motion for mesons are given by
\begin{eqnarray}
\label{eq:m1}
& & -\Delta\sigma + m_\sigma^2\sigma+g_2 \sigma^2+g_3 \sigma^3 =
-g_\sigma \rho_s -g^{\Lambda}_\sigma \rho^\Lambda_s ,
\\
\label{eq:m2}
& & -\Delta\omega + m_\omega^2\omega+c_3 \omega^3 =
 g_\omega \rho_v +g^\Lambda_\omega \rho^\Lambda_v
+\frac{f^{\Lambda}_{\omega}}{2M_{\Lambda}} \rho^\Lambda_T,
\\
\label{eq:m3}
& & -\Delta\rho + m_\rho^2\rho =
g_\rho \rho_3,
\\
\label{eq:m4}
& & -\Delta\sigma^* + m_{\sigma^*}^2\sigma^* =
-g^{\Lambda}_{\sigma^*} \rho^\Lambda_s,
\\
\label{eq:m5}
& & -\Delta\phi + m_\phi^2\phi =
g^\Lambda_\phi \rho^\Lambda_v,
\\
\label{eq:m6}
& & -\Delta A =
e \rho_p,
\end{eqnarray}
where $\rho_s$ ($\rho_s^\Lambda$), $\rho_v$ ($\rho_v^\Lambda$),
$\rho^\Lambda_T$, $\rho_3$, and $\rho_p$ are the scalar, vector,
tensor, third component of isovector, and proton densities, respectively.
The preceding coupled equations should be solved self-consistently
for various hypernuclear systems. We restrict our study to the spherical
case, and take into account the pairing contribution for open shell nuclei
by using the BCS theory.

There are several recipes for performing the center-of-mass correction
in the RMF theory, which have been discussed and compared in Ref~\cite{cm}.
It is well known that the relative contribution of the center-of-mass
correction to the total binding energy is very large for light nuclei,
therefore, we should treat the center-of-mass correction seriously.
In the present calculation, we take the microscopic
scheme suggested in Ref~\cite{cm},
\begin{eqnarray}
\label{eq:cm}
E_{\rm c.m.}=\frac{\left\langle F \right\vert \hat{\bf P}^{2}_{\rm total}
                   \left\vert   F \right\rangle} {2M_{\rm total}},
\end{eqnarray}
where
$M_{\rm{total}} = \displaystyle\sum M_B
                = n M_{\Lambda} + (A-n) M_N$
is the total mass of the system containing $n$ $\Lambda$ hyperons,
while
$\hat{\bf P}_{\rm{total}} = \displaystyle\sum \hat{\bf P}_B$
is the total momentum operator.  The expectation value of the square
of the total momentum operator is calculated from the actual wave
function of the ground state of the hypernucleus.
We perform self-consistent calculations for the
hypernuclear systems containing one or two $\Lambda$ hyperons.

\section{Parameters}
\label{sec:parameters}

For the parameters of the nucleonic sector,  we employ two successful
parameter sets TM1 and NL-SH as listed in Table~\ref{tab:tm1},
which could provide excellent descriptions for nuclear matter
and finite nuclei including unstable nuclei~\cite{tm1,nlsh}.
The TM1 set includes non-linear terms of both $\sigma$ and $\omega$ mesons,
while the NL-SH set contains only non-linear $\sigma$ terms.
As for the meson-hyperon couplings, it is well known that the properties
of single-$\Lambda$ hypernuclei are very sensitive to the ratios of the
meson-hyperon couplings to the meson-nucleon couplings,
    $R_\sigma=g^\Lambda_\sigma/g_\sigma$
and $R_\omega=g^\Lambda_\omega/g_\omega$~\cite{1lth6}.
We take the naive quark model value for the relative $\omega$ coupling
as $R_\omega=2/3$, while the relative $\sigma$ coupling is constrained by
fitting to the experimental $\Lambda$ binding energies of single-$\Lambda$
hypernuclei. It is shown in Fig.~1 that $R_\sigma=0.621$ could fairly reproduce
the experimental values of $\Lambda$ binding energies in a wide range of mass
number $A$ for both TM1 and NL-SH parameter sets. In the present calculation,
we adopt the quark model value of the tensor coupling,
$f^{\Lambda}_{\omega} = -g^{\Lambda}_{\omega}$~\cite{1lth2}.
The inclusion of tensor coupling is important to produce small spin-orbit
spitting of single-$\Lambda$ hypernuclei~\cite{1lth2,1lth3,1lth4}.

After the meson-hyperon couplings are determined by the experimental $\Lambda$
binding energies of single-$\Lambda$ hypernuclei, the RMF theory can be applied
straightforwardly to predict the properties of double-$\Lambda$ hypernuclei.
This is referred to as model 1 in the present calculation,
in which the exchanged mesons are limited to the usual mesons $\sigma$, $\omega$,
and $\rho$. There are some arguments that the two mesons $\sigma^*$ and $\phi$
may give essential contributions to multi-hyperon systems~\cite{prl93,2lth5}.
In this work, we incorporate these two additional mesons in model 2,
and investigate the dependence of the contributions on the coupling constants.
We take the experimental meson masses as $m_{\sigma^*}=980\;\rm{MeV}$ and
$m_{\phi}=1020\;\rm{MeV}$. For the $\phi$ coupling, we adopt the quark model
relationship $ R_\phi=g^\Lambda_{\phi} / g_{\omega} = -\sqrt{2}/3$, while the
relative $\sigma^*$ coupling, $R_{\sigma^*} = g^\Lambda_{\sigma^*} / g_{\sigma}$,
is taken as an adjustable parameter in model 2.
The two mesons $\sigma^*$ and $\phi$ were originally introduced to obtain strong
$\Lambda\Lambda$ attraction ($\triangle B_{\Lambda\Lambda}\approx 5\;\rm{MeV}$)
deduced from the earlier measurement~\cite{prl93}.
However, it is now believed that the $\Lambda\Lambda$ interaction is much weaker
($\triangle B_{\Lambda\Lambda}\approx 1\;\rm{MeV}$) as suggested by the striking
Nagara event. We first determine the relative coupling $R_{\sigma^*}$ by
reproducing the experimental value $\triangle B_{\Lambda\Lambda}\approx 1\;\rm{MeV}$
deduced from the Nagara event, then change the value of $R_{\sigma^*}$ in some range
to examine the dependence of the results on this parameter.

\section{Results}
\label{sec:results}

We calculate several double-$\Lambda$ hypernuclei including light, medium,
and heavy systems within the framework of RMF theory by using two successful
parameter sets TM1 and NL-SH as listed in Table~\ref{tab:tm1}.
For the meson-hyperon couplings, we adopt $R_\sigma=0.621$,
$R_\omega=2/3$, and $f^{\Lambda}_{\omega} = -g^{\Lambda}_{\omega}$,
which could give reasonable descriptions of single-$\Lambda$ hypernuclei
in a wide range of mass number $A$ as shown in Fig.~1.
We present in Table~\ref{tab:2lb} the $\Lambda\Lambda$ binding
energy $B_{\Lambda\Lambda}$, which is obtained for the hypernucleus
$^{A}_{\Lambda\Lambda}Z$ by
\begin{eqnarray}
\label{eq:2blambda}
B_{\Lambda\Lambda} \left (^{A}_{\Lambda\Lambda}Z \right)
            &=&  B\left (^{A}_{\Lambda\Lambda}Z \right) - B\left (^{A-2}Z \right) \\  \nonumber
            &=&  M\left (^{A-2}Z \right)
               - M\left (^{A}_{\Lambda\Lambda}Z \right) + 2 M_\Lambda .
\end{eqnarray}
We also list the quantity $\triangle B_{\Lambda\Lambda}$, which is called the
$\Lambda\Lambda$ bond energy, defined by
\begin{eqnarray}
\label{eq:2bdlambda}
\triangle B_{\Lambda\Lambda} \left (^{A}_{\Lambda\Lambda}Z \right)
=B_{\Lambda\Lambda}(^{A}_{\Lambda\Lambda}Z) - 2B_{\Lambda}(^{A-1}_{\Lambda}Z) .
\end{eqnarray}
The limited experimental data of light double-$\Lambda$ hypernuclei
are listed for comparison, while the calculated results of heavy systems
are considered as the predictions in the present models.
It is seen that $B_{\Lambda\Lambda}$ increases with increasing mass
number $A$, while $\triangle B_{\Lambda\Lambda}$ decreases.
We show in this table the calculated results with both TM1 and NL-SH
parameter sets by using models 1 and 2. In model 1, the exchanged mesons
are limited to the usual mesons $\sigma$, $\omega$, and $\rho$,
whose coupling constants are determined by the experimental data
of single-$\Lambda$ hypernuclei. Therefore, no more adjustable parameters
exist when model 1 is used to the calculation of double-$\Lambda$ hypernuclei.
In model 2, two additional mesons $\sigma^*$ and $\phi$ are included,
so we should determine their couplings properly. Here the quark model value
$R_\phi=-\sqrt{2}/3$ is adopted, while $R_{\sigma^*}=0.57$ (TM1) and
$R_{\sigma^*}=0.56$ (NL-SH) are constrained by the experimental value
$\triangle B_{\Lambda\Lambda} (^{6}_{\Lambda\Lambda}\rm{He}) \approx 1\;\rm{MeV}$
deduced from the Nagara event~\cite{2lexp5}. Without adjusting any parameter
in model 1, the calculated $\triangle B_{\Lambda\Lambda} (^{6}_{\Lambda\Lambda}\rm{He})$
is very close to the experimental value. It is found that the results of model 2
are almost identical to those of model 1. This is because the contributions
from $\sigma^*$ and $\phi$ mesons in model 2 are mostly cancelled with each
other, so there is no obvious difference between the two models for the
calculations of double-$\Lambda$ hypernuclei. However, there might be some
noticeable contributions from $\sigma^*$ and $\phi$ mesons when model 2 is
applied to the studies of multi-strange systems and neutron stars.

It is very interesting to discuss the influence of $\Lambda$ hyperons on the
nuclear core, which is known as the core polarization effect~\cite{1lth1,prc98s,2lth5}.
We compare the properties of the nucleus with those of the nuclear core
in single or double-$\Lambda$ hypernuclei. It is known that nucleons in the
core would be affected by the additional $\Lambda$ hyperons in hypernuclei.
The so-called rearrangement energy $E_R$ quantifies the core polarization effect,
which represents the change of nuclear core binding energies caused by the presence
of $\Lambda$, given by
\begin{eqnarray}
\label{eq:er}
E_R&=& \sum_{i=1}^{n} \epsilon_\Lambda^i - B_{n\Lambda}(^{A}_{n\Lambda}Z),
\end{eqnarray}
where $\epsilon_\Lambda^i$ is the absolute value of $\Lambda$ single-particle
energy, and $n$ denotes the number of $\Lambda$ hyperons in the hypernucleus.
We present in Table~\ref{tab:2lc} the calculated results of several
nuclei together with corresponding single and double-$\Lambda$ hypernuclei
in the RMF model with TM1 parameter set, and the results of double-$\Lambda$
hypernuclei are obtained by using model 1. It is seen that the rearrangement
energy $E_R$ decreases rapidly with increasing mass number $A$, and is usually
negligible in comparison with the binding energy except for very light systems.
For example, the rearrangement energy of $^{42}_{\Lambda\Lambda}\rm{Ca}$,
which contains two $1s$ $\Lambda$ hyperons coupled to the nuclear core
$^{40}\rm{Ca}$, is $0.29\;\rm{MeV}$, and this value is less than $1\%$ of
the $\Lambda\Lambda$ binding energy
($B_{\Lambda\Lambda} = 38.15 \;\rm{MeV}$, see Table~\ref{tab:2lb}).
But, the rearrangement energy of $^{6}_{\Lambda\Lambda}\rm{He}$
($E_R=3.54\;\rm{MeV}$) is rather large in comparison with the
$\Lambda\Lambda$ binding energy ($B_{\Lambda\Lambda} = 5.52 \;\rm{MeV}$).
We note that $B_{\Lambda\Lambda}=2\epsilon_\Lambda-E_R$ for double-$\Lambda$
hypernuclei. Therefore, the rearrangement energy for a light system may
significantly contribute to the binding energy, and is not negligible.
In Ref.~\citen{2lth5}, the rearrangement energy of the double-$\Lambda$ hypernucleus
has been assumed to be twice the one of the single-$\Lambda$ hypernucleus.
It is seen from Table~\ref{tab:2lc} that this assumption is satisfied
within an error of $\sim 10\%$ except for $^{210}_{\Lambda\Lambda}\rm{Pb}$
which has a negligible rearrangement energy.
On the other hand, the single-particle energies of neutrons and protons
in hypernuclei are also affected by the presence of $\Lambda$ hyperons.
We list the single-particle energies of neutrons at $1s$ states
($\epsilon_n(1s)$) and rms radii ($r_\Lambda$, $r_n$, and $r_p$)
in Table~\ref{tab:2lc}. It is shown that the change of the rms radii
of neutrons and protons is quite small. Usually the radius of a particle
increases as the strength of its potential decreases. The $\Lambda$ potential
is much shallower than the nucleon potential, so the rms radius of $\Lambda$
at $1s$ state should be rather larger than the one of nucleon at $1s$ state,
but it may be smaller than the radius of nucleon at higher state in heavy systems.
Hence, the rms radius of $\Lambda$ is larger than the total rms radius
of neutrons or protons in a light system, while it should be smaller than those
in heavy systems. It is obvious that the influence of $\Lambda$ on the nuclear core
is significantly weakened with the increase of mass number $A$.

For very light hypernuclei like $^{6}_{\Lambda\Lambda}\rm{He}$,
the center-of-mass correction provides a significant contribution
to the total binding energy, and there exist considerable
differences between different schemes to take into account the
center-of-mass correction. For instance, the value of the
center-of-mass correction of $^{6}_{\Lambda\Lambda}\rm{He}$
obtained from Eq.~(\ref{eq:cm}) is $10.25\;\rm{MeV}$ in model 1
with NL-SH parameter set, while it turns out to be
$16.92\;\rm{MeV}$ with the usual approximation
\begin{eqnarray}
\label{eq:cms}
E_{\rm c.m.}=\frac{3}{4} \cdot 41 \; A^{-1/3} \;\rm{(MeV)},
\end{eqnarray}
so it is rather large in comparison with the total binding energy
of $32.66\;\rm{MeV}$ obtained in this case.
The influence of the center-of-mass correction on
$\triangle B_{\Lambda\Lambda}$ is visible for light hypernuclei.
The simple approximation Eq.~(\ref{eq:cms}) yields
$\triangle B_{\Lambda\Lambda} (^{6}_{\Lambda\Lambda}\rm{He}) = 0.73 \;\rm{MeV}$
in agreement with the value presented in Table 2 for parametrization P5
of Ref.~\citen{2lth6}, while the microscopic scheme Eq.~(\ref{eq:cm}) results in
$\triangle B_{\Lambda\Lambda} (^{6}_{\Lambda\Lambda}\rm{He}) = 1.08\;\rm{MeV}$
as shown in Table~\ref{tab:2lb}. Therefore, we emphasize that
the center-of-mass correction plays an important role in light systems,
but its influences on hypernuclear properties and the differences between
different schemes decrease rapidly with increasing mass number $A$.
In this work, the calculated results of $B_{\Lambda}(^{5}_{\Lambda}\rm{He})$
and $B_{\Lambda\Lambda} (^{6}_{\Lambda\Lambda}\rm{He})$
are underestimated in comparison with the experimental values.
This might be due to the limited reliability of the RMF approach
for light systems.

So far it is difficult to determine a reliable $\sigma^*$ coupling by scarce
and various difficult measurements of double-$\Lambda$ hypernuclei.
In order to investigate the possible contributions from the two mesons
$\sigma^*$ and $\phi$, we plot $\triangle B_{\Lambda\Lambda}$ of several
double-$\Lambda$ hypernuclei as a function of $R_{\sigma^*}$ in Fig.~2.
Some large values of $R_{\sigma^*}$ were used in previous studies~\cite{prl93,2lth5}
in order to obtain strong $\Lambda\Lambda$ interaction as
$\triangle B_{\Lambda\Lambda} \approx 5\;\rm{MeV}$ deduced from the earlier
measurement~\cite{2lexp1,2lexp2}. This has been changed by the Nagara
event~\cite{2lexp5}, which leads to a small $\Lambda\Lambda$ bond energy
$\triangle B_{\Lambda\Lambda} \approx 1\;\rm{MeV}$.
It is found that the contributions from $\sigma^*$ and $\phi$
for heavy hypernuclei are much smaller than those for light systems.
For instance, at $R_{\sigma^*} \approx 0.75$ the light hypernuclei
$^{6}_{\Lambda\Lambda}\rm{He}$ gets about $4\;\rm{MeV}$ extra binding energy
in model 2 compared with the result in model 1, while it is less than
$1\;\rm{MeV}$ for heavy hypernuclei like $^{210}_{\Lambda\Lambda}\rm{Pb}$.
For each species of the double-$\Lambda$ hypernuclei,
$\triangle B_{\Lambda\Lambda}$ shows a decrease tendency
with decreasing $R_{\sigma^*}$.

There are many discussions in the literature about possible mechanisms
to soften the equation of state (EOS) of neutron star matter,
e.q., by hyperons, kaon condensates, or even quark
phases~\cite{prc96s,prc99,jpg02s,prc04s,pr00,prc02}.
It is well known that the hyperon-hyperon interactions play important
roles in determining the abundance of hyperons at high densities.
The Nagara event~\cite{2lexp5} has suggested a much weaker $\Lambda\Lambda$
interaction than the one used in early calculations~\cite{prc96s}.
We would like to discuss the neutron star properties changed
due to the weak $\Lambda\Lambda$ interaction.
Here we focus on the contribution of $\Lambda$ hyperons only,
since we have rather poor knowledge of other hyperon interactions.
It is shown in Ref.~\citen{1lth3} that the inclusion of $\Lambda$
hyperons could considerably soften the EOS at high densities.
We compare the results of neutron star properties, which are calculated
in model 2 with TM1 parameter set. The maximum mass of neutron stars
using $R_{\sigma^*}=0.57$, which is constrained by $\triangle B_{\Lambda\Lambda}
\approx 1\;\rm{MeV}$, is about $0.1\;M_\odot$ larger than the one using
$R_{\sigma^*}=0.75$ determined by $\triangle B_{\Lambda\Lambda}
\approx 5\;\rm{MeV}$. It is obvious that the weak $\Lambda\Lambda$ interaction
suggested by the Nagara event leads to a slightly stiffer EOS than the
strong $\Lambda\Lambda$ interaction.

\section{Summary}
\label{sec:summary}

In summary, we have performed the self-consistent calculations
for double-$\Lambda$ hypernuclei in a wide range of mass
number $A$ within the framework of RMF theory.
We have adopted the parameter sets TM1 and NL-SH, which could provide
excellent descriptions for finite nuclei and single-$\Lambda$ hypernuclei.
We have studied the properties of double-$\Lambda$ hypernuclei using two models.
Model 1 is the RMF theory containing the exchanged mesons $\sigma$, $\omega$,
and $\rho$. Model 2 incorporates two additional  mesons $\sigma^*$
and $\phi$, which couple exclusively to hyperons.
The results of model 2 are definitely dependent on the couplings of $\sigma^*$
and $\phi$. With the couplings constrained by
$\triangle B_{\Lambda\Lambda}\approx 1\;\rm{MeV}$ deduced from the Nagara event,
the results of model 2 are almost identical to those of model 1.

The influence of $\Lambda$ hyperons on the nuclear core has been investigated
by comparing various properties of the nucleus with those of the nuclear core
in single or double-$\Lambda$ hypernuclei. We found that the rearrangement
energy decreases rapidly with increasing mass number $A$, and could be
neglected in most cases. But for very light system like $^{6}_{\Lambda\Lambda}\rm{He}$,
the rearrangement energy may be rather large in comparison with the
$\Lambda\Lambda$ binding energy, and is not negligible.
The single-particle energies of neutrons in hypernuclei are also affected
by the presence of $\Lambda$ hyperons.

We have examined the possible contributions from the two strange mesons
$\sigma^*$ and $\phi$. They could give either positive or negative
contributions to the $\Lambda\Lambda$ binding energies, which depend
on the coupling constants used in the calculation.
It is found that the contributions from $\sigma^*$ and $\phi$ to
$B_{\Lambda\Lambda}$ for heavy hypernuclei are much smaller than
those for light systems. So far it is difficult to determine a reliable
$\sigma^*$ coupling by the scarce and various difficult measurements
of double-$\Lambda$ hypernuclei. It is necessary and important to get
more and better experimental data on double-$\Lambda$ hypernuclei
so that theoretical models can be checked and extended to further study.

\section*{Acknowledgments}
This work was supported in part by the National Natural Science
Foundation of China (No. 10135030) and the Specialized Research Fund
for the Doctoral Program of Higher Education (No. 20040055010).

%


\vspace*{1.0cm}
\begin{table}
\caption{The parameter sets TM1~\cite{tm1} and NL-SH~\cite{nlsh}. The masses are given in $\rm{MeV}$.}
\vspace{0.0cm}
\begin{center}
\begin{tabular}{ccccccccccc} \hline
\hline
   & $M$    & $m_\sigma$ & $m_\omega$ & $m_\rho$ & $g_\sigma$ & $g_\omega$ & $g_\rho$
   & $g_2$ ($\rm{fm}^{-1}$) & $g_3$  & $c_3$
\\ \hline
TM1   & 938.0 & 511.198 & 783.0 & 770.0 & 10.0289 & 12.6139 & 4.6322 & -7.2325 & 0.6183 & 71.3075 \\
NL-SH & 939.0 & 526.059 & 783.0 & 763.0 & 10.444  & 12.945  & 4.383  & -6.9099 &-15.8337& 0.0     \\
\hline
\hline
\end{tabular}
\end{center}
\label{tab:tm1}
\end{table}

\vspace*{2.0cm}
\begin{table}
\caption{$B_{\Lambda\Lambda}$ and $\triangle B_{\Lambda\Lambda}$ of double-$\Lambda$
    hypernuclei. The calculated results of models 1 and 2 are denoted by 1 and 2, respectively.
    The available experimental data are taken from Refs.~\citen{2lexp1,2lexp2,2lexp3,2lexp4,2lexp5}.}
\begin{tabular}{lrrrrrrrrrr}\hline
\hline
   & $B_{\Lambda\Lambda}$&  &  TM1 &      & NL-SH &  $\triangle B_{\Lambda\Lambda}$ & & TM1 & & NL-SH  \\
\cline{3-4} \cline{5-6} \cline{8-9} \cline{10-11}  \vspace{-0.2cm}\\
                                 &       exp.      & 1      & 2    &  1    & 2     &   exp.        & 1     & 2     & 1     & 2      \\
\hline
  $^{6}_{\Lambda\Lambda}{\rm He}$& $7.25 \pm 0.2 $ &  5.52  &  5.48& 4.75  &  4.68 & $1.0 \pm 0.2$ &  1.07 &  1.03 &  1.08 &  1.01  \\\vspace{-0.1cm}
 $^{10}_{\Lambda\Lambda}{\rm Be}$& $17.7 \pm 0.4 $ & 16.34  & 16.28& 16.03 & 15.94 & $4.3 \pm 0.4$ &  0.37 &  0.31 &  0.38 &  0.29  \\\vspace{-0.1cm}
                                 & $14.6 \pm 0.4 $ &        &      &       &       & $1.2 \pm 0.4$ &       &       &       &        \\
                                 & $ 8.5 \pm 0.7 $ &        &      &       &       & $-4.9\pm 0.7$ &       &       &       &        \\
 $^{13}_{\Lambda\Lambda}{\rm B }$& $27.5 \pm 0.7 $ & 22.14  & 22.07& 22.65 & 22.52 & $4.8 \pm 0.7$ &  0.26 &  0.19 &  0.33 &  0.21  \\
 $^{18}_{\Lambda\Lambda}{\rm O }$&                 & 25.89  & 25.85& 25.30 & 25.23 &               &  0.14 &  0.10 &  0.14 &  0.07  \\
 $^{42}_{\Lambda\Lambda}{\rm Ca}$&                 & 38.15  & 38.13& 37.90 & 37.86 &               &  0.04 &  0.02 &  0.04 &  0.00  \\
 $^{92}_{\Lambda\Lambda}{\rm Zr}$&                 & 47.11  & 47.10& 47.73 & 47.71 &               &  0.03 &  0.02 &  0.04 &  0.02  \\
$^{210}_{\Lambda\Lambda}{\rm Pb}$&                 & 52.19  & 52.19& 53.03 & 53.02 &               &  0.03 &  0.02 &  0.02 &  0.02  \\
\hline
\hline
\end{tabular}
\label{tab:2lb}
\end{table}

\vspace*{3.0cm}
\begin{table}
\caption{Comparison of the energies (in $\rm{MeV}$) and radii (in $\rm{fm}$) of single and double-$\Lambda$ hypernuclei
         with those of normal nuclei. The results are obtained with TM1 parameter set, and the model 1 is used for
         the calculation of double-$\Lambda$ hypernuclei. $B_{\rm{total}}$ and $B_{\rm{core}}$ represent the total
         binding energies and the binding energies of the nuclear core. $E_R$ is the rearrangement energy given
         in Eq.~(\ref{eq:er}). $\epsilon_\Lambda(1s)$ and $\epsilon_n(1s)$ are the absolute values of single-particle
         energies for $\Lambda$ and neutron at $1s$ states. The rms radii of $\Lambda$, neutron, and proton are
         denoted by $r_\Lambda$, $r_n$, and $r_p$, respectively. }
\begin{tabular}{lllrrrrrrrrrl}
\hline
\hline
& & & $B_{\rm{total}}$ & $B_{\rm{core}}$ & $E_{\rm{R}}$ & $\epsilon_\Lambda(1s)$ & $\epsilon_n(1s)$ & & $r_\Lambda$ & $r_n$ &  $r_p$ & \\
\hline
&                  $^{ 4}{\rm He}$& &  28.19 &  28.19 &        &        & 17.75  & &      & 1.90 & 1.92 &  \\
&        $^{ 5}_{\Lambda}{\rm He}$& &  30.42 &  26.32 &  1.87  &  4.10  & 19.35  & & 2.77 & 1.90 & 1.91 &  \\
& $^{ 6}_{\Lambda\Lambda}{\rm He}$& &  33.71 &  24.65 &  3.54  &  4.53  & 20.90  & & 2.74 & 1.89 & 1.90 &  \\
\hline
&                  $^{16}{\rm O }$& & 128.73 & 128.73 &        &        & 40.67  & &      & 2.56 & 2.58 &  \\
&        $^{17}_{\Lambda}{\rm O }$& & 141.61 & 128.36 &  0.37  & 13.24  & 41.42  & & 2.48 & 2.56 & 2.59 &  \\
& $^{18}_{\Lambda\Lambda}{\rm O }$& & 154.62 & 127.90 &  0.83  & 13.36  & 42.18  & & 2.49 & 2.57 & 2.59 &  \\
\hline
&                  $^{40}{\rm Ca}$& & 344.35 & 344.35 &        &        & 51.58  & &      & 3.32 & 3.36 &  \\
&        $^{41}_{\Lambda}{\rm Ca}$& & 363.40 & 344.22 &  0.13  & 19.18  & 52.04  & & 2.76 & 3.32 & 3.36 &  \\
& $^{42}_{\Lambda\Lambda}{\rm Ca}$& & 382.49 & 344.06 &  0.29  & 19.22  & 52.51  & & 2.77 & 3.32 & 3.37 &  \\
\hline
&                 $^{208}{\rm Pb}$& &1638.48 &1638.48 &        &        & 55.89  & &      & 5.75 & 5.48 &  \\
&       $^{209}_{\Lambda}{\rm Pb}$& &1664.56 &1638.45 &  0.03  & 26.11  & 56.07  & & 4.09 & 5.75 & 5.48 &  \\
&$^{210}_{\Lambda\Lambda}{\rm Pb}$& &1690.67 &1638.39 &  0.09  & 26.14  & 56.26  & & 4.09 & 5.75 & 5.48 &  \\
\hline
\hline
\end{tabular}
\label{tab:2lc}
\end{table}

\vspace*{4.0cm}
\begin{figure}[h]
\centerline{\includegraphics[bb=65 430 520 740, width=10 cm, clip]{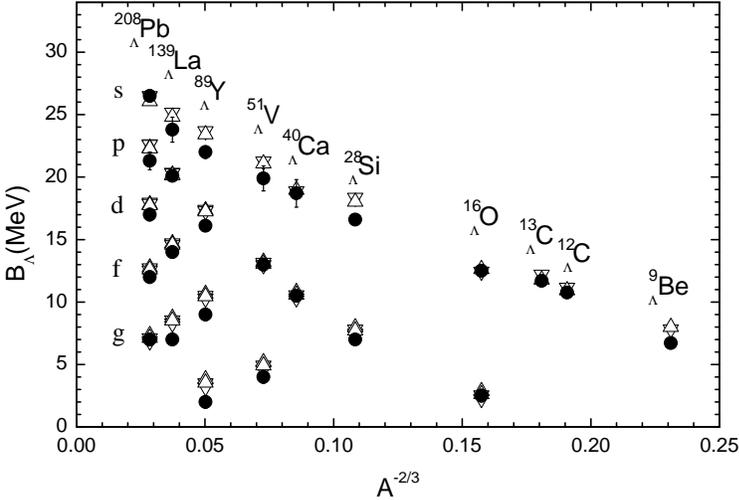}}
\caption{The $\Lambda$ binding energies in single-$\Lambda$
    hypernuclei. The solid circles represent the experimental
    data with errors taken from Refs.~\protect\citen{npa88,prl91,prc96}.
    The open up and down triangles are the results in the RMF
    model with the parameter sets TM1 and NL-SH, respectively. }
\end{figure}

\vspace{1.0cm}
\begin{figure}[h]
\centerline{\includegraphics[bb=50 310 510 750, width=10 cm, clip]{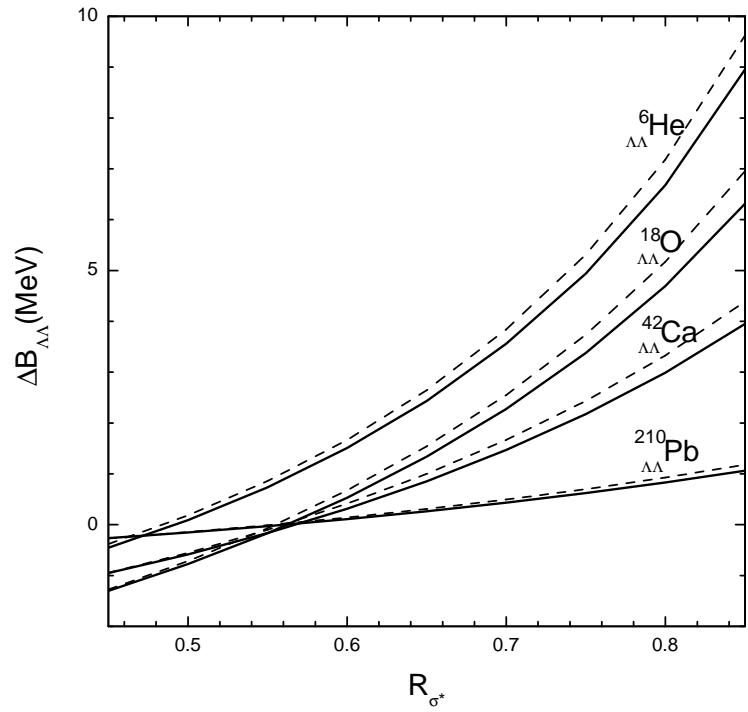}}
\caption{$\triangle B_{\Lambda\Lambda}$ as a function of $R_{\sigma^*}$
    for several double-$\Lambda$ hypernuclei.
    The solid lines represent the results in the RMF model
    with TM1 parameter set, while those with NL-SH set are shown
    by dashed curves.}
\end{figure}


\end{document}